\documentclass[pra,preprint,amsmath,amssymb,showpacs]{revtex4}
\usepackage{bm}
\usepackage{graphics}

\begin{document}

\title{Stationary states in a system of two linearly coupled 2D NLS equations
with nonlinearities of opposite signs}

\author{Valery S. Shchesnovich }
\email{valery@loqnl.ufal.br}
\author{Solange B. Cavalcanti}
\email{solange@loqnl.ufal.br} \affiliation{ Departamento de
F\'{\i}sica - Universidade Federal de Alagoas, Macei\'o 7072-970,
Brazil }

\begin{abstract}

We study, analytically and numerically, the stationary states in
the system of two linearly coupled nonlinear Schr{\"o}dinger
equations in two spatial dimensions, with the nonlinear
interaction coefficients of opposite signs. This system is the
two-dimensional analog of the coupled-mode equations for a
condensate in the double-well trap  [\textit{Physical Review A}
\textbf{69}, 033609 (2004)].  In contrast to the one-dimensional
case, where the bifurcation from zero leads to stable bright
solitons, in two spatial dimensions this bifurcation results in
the appearance of unstable soliton solutions (the Townes-type
solitons). With the use of a parabolic potential the ground state
of the system is stabilized. It corresponds to strongly coupled
condensates and is stable with respect to collapse. This is in
sharp contrast to the one-dimensional case, where the ground state
corresponds to weakly coupled condensates and is unstable.
Moreover, the  total number of atoms of the stable ground state
can be much higher than the collapse threshold for a single
two-dimensional condensate with a negative $s$-wave scattering
length.
\end{abstract}

 \maketitle

\section{Introduction}
\label{sec1}

Bose-Einstein condensates (BECs) in trapped dilute gases exhibit
interesting interplay between  quantum coherence and nonlinearity
since, at zero temperature, the quantum gas is described by the
mean-field theory based on the nonlinear Schr\"odinger equation
with an external potential -- the Gross-Pitaevskii (GP) equation
\cite{GPE} for the order parameter. The macroscopic quantum
coherence of BEC, first experimentally demonstrated in Refs.
\cite{intrfexp,intrf2comp}, was subsequently explained
theoretically  \cite{intrfth} with the use of the GP equation.

Nonlinear phenomena in BEC bear similarity with  nonlinear optics.
Similar to optics, where the bright and dark solitons are
supported respectively by the focusing and defocusing
nonlinearities, in BECs the $s$-wave scattering length is the
determining factor. Dark solitons are routinely observed in the in
the quasi-one-dimensional condensates with repulsive interactions
\cite{dark1,dark2,dark3,dark4}. On the other hand, the attractive
one-dimensional BEC propagates in the form of the bright solitons
\cite{bright1,bright2}.

The $s$-wave scattering length can be modified by application of
magnetic field near the Feshbach resonance \cite{fesh}. For a BEC
constrained to lower spatial dimensions, the Feshbach resonance
still proves to be sharp, for instance, in the two-dimensional
condensate \cite{Feshb2D}. The feasibility of control over the
scattering length in BEC by the optical means was also proposed
\cite{optcontr} (see also Refs.~\cite{opt1,opt2}).

Control over the scattering length in a part of the condensate can
be realized in the double-well trap with far-separated wells.
Condensates in the double-well potential are currently routinely
created and studied  in the experiments (see, for instance,
Refs.~\cite{DW1exp,DW2exp,DWexp3}).

The double-well trap is created in one spatial dimension, the
other two dimensions thus allow for two geometrically distinct
cases which correspond to the one- and two-dimensional BECs
depending on the trap asymmetry. A combination of the double-well
trap with the control over the scattering length allows one to
observe two tunnel-coupled BECs with the opposite interactions
(i.e., one attractive and the other repulsive). In one spatial
dimension such a setup leads to appearance of the unusual stable
bright solitons \cite{PRA,PHYSD} which have almost all atoms
contained in the repulsive condensate. The one-dimensional ground
state, however, corresponds to weakly coupled condensates and is
unstable with respect to collapse \cite{PRA}. In this paper we
study the two-dimensional case. We find that the two-dimensional
solitons (the Townes type solitons) are always unstable. However,
with the help of a confining potential, the ground state can be
stabilized. In contrast to the one-dimensional case, in the
two-dimensions the ground state of the system corresponds to
strongly-coupled condensates and there is an energy barrier for
collapse  (see also Ref. \cite{EnerBarr}.)

The applicability of the GP-based mean-field theory is limited,
but it always applies to the description of stable stationary
states. This is due to the similarity between the Bogoliubov-de
Gennes equations, describing the surrounding cloud of hot atoms,
and the equations describing evolution of a linear perturbation of
the order parameter \cite{APPL}. Thus, a stable mean-field
stationary state is also stable in the full quantum approach, its
life time is equal to that of the condensate.

Theoretical investigations of BEC in the double-well trap go back
to the prediction of the anomalous Josephson oscillations
\cite{AJO} and the macroscopic quantum self-trapping of the
condensate \cite{cmBEC1,cmBEC2,cmBEC3}.

Experimental advances \cite{DW1exp,DW2exp,DWexp3} in the
production  and manipulation of the condensate in the double-well
trap make  the implementation of the tunnel-coupled repulsive and
attractive condensates feasible. Recently a direct observation of
the quantum tunneling and nonlinear self-trapping of a BEC in the
double-well trap was reported \cite{DWexp3}.

The collapse instability in an attractive BEC is loaded into the
double-well trap was also recently investigated. In the quasi-one
dimensional case, the critical number for collapse turns out to be
larger than the same for the corresponding axially-symmetric
harmonic trap \cite{CRNUM}. The tunnelling-induced collapse of an
attractive BEC in a double-well trap can take place  under the
influence of a time-dependent potential \cite{TIC}.

Control over the scattering length in one part of a condensate can
be realized in a double-well trap with far separated wells. In
this case a simplification of the GP equation is possible, which
results in the coupled-mode approximation similar to that of Refs.
\cite{cmBEC2,cmBEC3}. However, in contrast to the latter works,
the kinetic energy of the condensate is taken into account (see
section \ref{sec2}).  The coupled-mode system also has
applications in optics  (see, for instance, Refs.
\cite{cmOP1,cmOP2,cmOP3}).

Since the two-dimensional (2D) nonlinear Schr\"odinger equation
(NLS),
\begin{equation}
i\partial_t\Psi + \nabla^2\Psi - g|\Psi|^2\Psi = 0,
 \label{EQ1}\end{equation}
is critical, we adopt the point of view based on the analysis of
the critical scaling and its perturbations. The NLS equation has
the following scale invariance: if $\Psi(t,\vec{r})$ is a solution
then $\widetilde{\Psi}(t,\vec{r}) = k\Psi(k^2t,k\vec{r})$ is also
a solution. The number of atoms $N$ (or the $l_2$-norm), defined
by $N = \int\mathrm{d}^n\vec{r}|\Psi|^2$,  is scaled as $\tilde{N}
= k^{2-n}N$ in $n$ spatial dimensions.

The  scale invariance of the 2D NLS equation allows for a family
of solutions with the same number of particles. One may call the
2D scale invariance the ``critical scaling''. The critical scaling
leads to important consequences  (see, for instance, Ref.
\cite{Collapse2D}). As the number of particles is constant for the
whole family of solutions, the Vakhitov-Kolokolov (VK) criterion
\cite{VK} applied to the Townes soliton gives marginal stability
(or instability), since $\partial N/\partial \mu = 0$, where $\mu$
is the chemical potential for a particular solution of the family
(i.e., $-\mu$ is the frequency).

This explains why addition of an external {\it confining }
potential allows for  stable localized solutions. Indeed, the
external potential  breaks the scale invariance and the number of
particle degeneracy is broken too: some solutions of the former
Townes soliton family have number of particles below the collapse
threshold (moreover, thanks to the so-called lens transformation
\cite{LENS}, the collapse threshold does not depend on the
strength of the potential if the latter is parabolic).

Besides addition of an external potential, there are other ways to
break the  scale invariance.  The coupled-mode system, i.e. the
system of  two linearly coupled NLS equations, provides another
way. The coupling coefficient is proportional to the tunnelling
rate through the central barrier.

Having understood the relation between the broken scale invariance
and stability against the collapse, one may wonder if the linear
coupling of two 2D NLS equations  allows for the existence of
stable two-dimensional (i.e. Townes-type) solitons.  Indeed, if
the critical scaling is broken and the number of atoms (or the
number of particles, generally) depends on the chemical potential
then there could be  self-localized stationary solutions (which
have right to be called solitons) with $\partial N/\partial\mu
<0$, i.e. satisfying the VK criterion, with the hope that they are
stable.

In the following, using the singular perturbation theory
supplemented by numerical simulations,  we argue that the
two-dimensional soliton solutions to the coupled-mode system are
always unstable with respect to collapse. The reason is that, in
contrast to the 1D case \cite{PRA},  the bifurcation from zero in
the 2D coupled-mode system is always discontinuous: the number of
atoms of the two-dimensional soliton solution with vanishing
amplitude does \textit{not} vanish. However,  addition of an
external confining potential (a potential transverse to the
double-well trap) restores the continuity of the bifurcation from
zero and leads to the appearance of stable solutions. Some of them
have the number of atoms much larger than the collapse threshold
in a single 2D NLS equation. Moreover, such a solution is the
ground state of the system, since it has the lowest possible
energy for a fixed number of atoms. In contrast to the
one-dimensional case, the ground state in the two-dimensional
system is secured from collapse by an energy barrier.

The paper is organized as follows. In section \ref{sec2} we derive
the coupled-mode system from the Gross-Pitaevskii equation for a
condensate in  an asymmetric double-well trap. In section
\ref{sec3} we consider the stability properties of the axially
symmetric stationary states. Section \ref{bigsec4} is devoted to
the study of the limiting case solutions: the bifurcation from
zero in the coupled-mode system with and without the parabolic
potential, sections \ref{sec4} and \ref{sec5}, respectively, and
the asymptotic solution corresponding to large negative values of
the chemical potential, section \ref{sec6}. The reader not
interested in details of the bifurcation analysis may go directly
to section \ref{sec7}, where we summarize the results and discuss
the numerical solution of the coupled-mode system. In the numerics
we have used the Fourier spectral collocation method and looked
for the stationary solutions using the numerical schemes of Refs.
\cite{PHYSD} and \cite{Rayleigh}. Section \ref{sec7} contains some
concluding remarks.

\section{Derivation of the coupled-mode system }
\label{sec2}

The coupled-mode system follows from the GP equation  under two
conditions. First, the wells of the double-well trap must be far
separated.  Second, the number of BEC atoms must be below a
certain threshold (equation (\ref{EQ12}) below) with the result
that the motion of the condensate in the spatial dimension of the
double-well trap is equivalent to that of a quantum particle.  The
stationary states considered below satisfy this condition.   It
follows that BEC atoms occupy only the degenerate energy levels of
the double-well trap.

The tunnelling coefficient, usually defined through an integral
over the overlap of the wave functions (see, for instance, Ref.
\cite{cmBEC2,cmBEC3}), can be, in fact, explicitly given in terms
of the parameters of the double-well trap (equation (\ref{EQ9})).
This is due to the above conditions and the simple fact that the
wave functions of the localized basis (given by equations
(\ref{EQ6}) and (\ref{EQ7})) are uniquely defined by the trap.

The two-dimensional coupled-mode system describes the so-called
pancake condensates. Although the 1D and 2D coupled-mode systems
are similar, there is a difference in the order of magnitude of
the respective parameters which is explained below.

The Gross-Pitaevskii equation for the order parameter
$\Psi(\vec{r},t)$ of a BEC in a double-well trap given by a
parabolic potential with a Gaussian barrier reads
\begin{equation}
i\hbar \partial_t\Psi = \left\{-\frac{\hbar^2}{2m} \nabla^2 +
V_\mathrm{ext}(\vec{r}) + g|\Psi|^2\right\}\Psi,
 \label{EQ3}\end{equation}
where
\begin{equation}
V_\mathrm{ext} = \frac{m}{2}(\omega_\perp^2\vec{r}_\perp{}^2 +
 \omega_z^2z^2)+ V_B\exp\left\{-\frac{(z-z_0)^2}{2\sigma^2}\right\},
\quad V_B>0.
 \label{EQ4}\end{equation}
Here the parameters $z_0$ and $\sigma$ give the position and width
of the  barrier, respectively. The pancake geometry corresponds to
the condition $\gamma\equiv \omega_z/\omega_\perp\gg1$. In terms
of the potential from equation (\ref{EQ4})  the wells of the
double well trap are far separated if $\sigma \sim \ell_z$ with
$\ell_z \equiv \sqrt{\frac{\hbar}{m\omega_z}}$ being the
oscillator length in the $z$-direction.

The  motion in the double-well trap  and the transverse dynamics
of BEC can be factorized if the number of BEC atoms is not large
(see equation (\ref{EQ12})), i.e. if the oscillator length
$\ell_z$ is much smaller than the characteristic length of
nonlinearity $\ell_\mathrm{nl}$.  In the case of a double-well
with far separated wells, the first two energy levels are
quasi-degenerate:
\begin{equation}
E_1-E_0\ll E_2-E_1.
 \label{EQ5}\end{equation}
Being interested in the ground state of the system (occupied by
the condensate)  we consider only the degenerate subspace. This
approximation neglects the uncondensed atomic cloud  which can be
discarded for temperatures below the condensation threshold
\cite{GPE}.

The basis in the degenerate subspace for expansion of the order
parameter is dictated by the necessity to simplify the nonlinear
term, which is not small for the transverse degrees of freedom,
i.e. in the pancake plane. Evidently, one should select such a
basis, say $\psi_u(z)$ and $\psi_v(z)$, where each basis function
is \textit{localized in just one of the wells}. The localized
basis is given by a rotation of the wave functions for the ground
state and the first excited state:
\begin{equation}
\psi_u(z) =  \frac{\psi_0(z) + \varkappa\psi_1(z)}{\sqrt{1 +
\varkappa^2}},\quad \psi_v(z) = \frac{\varkappa\psi_0(z)-
\psi_1(z)}{\sqrt{1 + \varkappa^2}}.
  \label{EQ6}\end{equation}
Obviously, the new wave functions are orthogonal and normalized.
The parameter $\varkappa$ is selected by the quotient of the
absolute values of the eigenfunctions $\psi_0(z)$ and $\psi_1(z)$
at the minima of the double-well potential $z_-$ and $z_+$ (say,
$z_-<z_+$) \cite{PRA}. We set
\begin{equation}
\varkappa = \frac{\psi_1(z_-)}{\psi_0(z_-)} \approx
-\frac{\psi_0(z_+)}{\psi_1(z_+)}
 \label{EQ7}\end{equation}
 (the positions of the extremals of the wave functions
slightly deviate from  the minima of the trap; these deviations we
neglect). For this choice of $\varkappa$ the wave functions
$\psi_u(z)$ and $\psi_v(z)$ defined by equation (\ref{EQ6}) are
localized in the left and right well, respectively (see also
figures 1 and 2 in Ref. \cite{PRA}).

The Hamiltonian for  a quantum particle in the double-well
potential, projected on the degenerate subspace, i.e. $H_z =
E_0|\psi_0\rangle \langle\psi_0| + E_1|\psi_1\rangle
\langle\psi_1|$, in the new basis becomes
\begin{equation}
H_z = \bar{E}(|\psi_u\rangle \langle\psi_u| +
|\psi_v\rangle\langle\psi_v|) +
|\psi_v\rangle\mathcal{E}\langle\psi_v| - K(|\psi_u\rangle
\langle\psi_v| + |\psi_v\rangle\langle\psi_u|).
 \label{EQ8}\end{equation}
Here
\begin{equation}
\bar{E} = \frac{E_0 + E_1}{2},\quad \mathcal{E} =
\frac{1-\varkappa^2}{1+\varkappa^2}(E_1 - E_0), \quad K =
\frac{\varkappa(E_1 - E_0)}{1+\varkappa^2}.
 \label{EQ9}\end{equation}
We can set $\bar{E} = 0$ without loss of generality. Equation
(\ref{EQ9}) gives the tunnelling coefficient $K$ and the
zero-point energy difference $\mathcal{E}$.

We approximate the solution of the GP equation (\ref{EQ3}) by a
sum of the factorized  order parameters:
\begin{equation}
\Psi(t,\vec{r}_\perp,z) = \Phi_u(t,\vec{r}_\perp)\psi_u(z) +
\Phi_v(t,\vec{r}_\perp)\psi_v(z).
 \label{EQ10}\end{equation}
Let us now formulate the condition for the factorization in
equation (\ref{EQ10}). We have neglected the nonlinear term in the
Gross-Pitaevskii equation as compared to the longitudinal kinetic
term, that is
\begin{equation}
\frac{\hbar^2}{2m}\frac{1}{\ell_z^2}  \gg |g||\Phi|^2|\psi|^2,
 \label{EQ11}\end{equation}
(for each of the two wells). The wave functions can be estimated
as follows $|\psi|^2 \sim 1/\ell_z$ and $|\Phi|^2 \sim
\mathcal{N}/d_\perp^2$, with $d_\perp$ being the transverse radius
of the condensate and $\mathcal{N}$ the number of atoms (in the
considered well). Using the expression for the nonlinear
coefficient in the Gross-Pitaevskii equation $g = 4\pi \hbar^2
a_s/m$ \cite{GPE}, where $a_s$ is the atomic scattering length, we
get the applicability condition for the separation of variables as
follows
\begin{equation}
\frac{8\pi |a_s| \ell_z \mathcal{N}}{d^2_\perp}\ll 1.
 \label{EQ12}\end{equation}
Condition (\ref{EQ12})  must be satisfied by both condensates in
the double-well trap.

The coupled-mode system is derived  by inserting  expansion
(\ref{EQ10}) into equation   (\ref{EQ3}), using the orthogonality
of the basis functions $\psi_{u,v}$, the formulae
\[
H_z\psi_u = -K\psi_v,\quad H_z \psi_v = \mathcal{E}\psi_v -
K\psi_u,
\]
and throwing away the small nonlinear terms involving the
cross-products of the localized wave functions $\psi_u$ and
$\psi_v$ (see also the discussion of Ref. \cite{PRA}). One arrives
at the system:
\begin{subequations}
\label{EQ13}
\begin{equation}
i\hbar \partial_t\Phi_u = -\frac{\hbar^2}{2m}\nabla_\perp^2\Phi_u
+ V(\vec{r}_\perp)\Phi_u + g_u|\Phi_u|^2 \Phi_u - K \Phi_v,
\label{EQ13a}\end{equation}
\begin{equation}
i\hbar \partial_t\Phi_v = -\frac{\hbar^2}{2m}\nabla_\perp^2\Phi_v
+ V(\vec{r}_\perp)\Phi_v + (\mathcal{E} + g_v|\Phi_v|^2) \Phi_v -
K \Phi_u. \label{EQ13b}\end{equation}
\end{subequations}
Here $g_{u,v} \equiv \int\mathrm{d}z g(z)|\psi_{u,v}|^4$ and
$V(\vec{r}_\perp) = \frac{m\omega_\perp^2}{2}\vec{r}_\perp{}^2$
(by changing the sign of either $\psi_u$ or $\psi_v$ one can
always set $K>0$).

System (\ref{EQ13}) is the basis of our approach. Conditions
(\ref{EQ5}) and (\ref{EQ12}) are satisfied by all stable
stationary states considered below. In the search for
two-dimensional soliton solutions we will  neglect the confining
transverse potential $V(\vec{r}_\perp)$ (i.e. when it is
considered as flat on the scale of the Townes-like soliton
solution). This case will be called below ``the coupled-mode
system without the transverse potential''. The atomic interaction
in the $u$-condensate is attractive, $g_u<0$, while in the
$v$-condensate it is externally modified to repulsive, $g_v>0$.

For the numerical and analytical analysis it is convenient to use
the dimensionless variables defined as follows
\begin{equation}
T = \frac{\omega_\perp}{2}t,\quad \vec{\rho} =
\frac{\vec{r}}{\ell_\perp},\quad \ell_\perp \equiv
\sqrt{\frac{\hbar}{m\omega_\perp}}.
 \label{EQ14}\end{equation}
The order parameters are expressed as
\begin{equation}
\Phi_u = \frac{\sqrt{\Delta}}{\ell_\perp}\,u,\quad \Phi_v =
\frac{\sqrt{\Delta}}{\ell_\perp}\, v,
 \label{EQ15}\end{equation}
where $\Delta =
\left(\sqrt{8\pi|a^{(u)}_s|}\int\mathrm{d}z|\psi_u|^4\right)^{-1}$
and $a^{(u)}_s$ is the scattering length in the $u$-condensate.
The dimensionless system reads
\begin{subequations}
\label{EQ16}
\begin{equation}
i\partial_T u = \left(-\nabla^2_{\vec{\rho}} +
\vec{\rho}\,{}^2\right)u - |u|^2u - \kappa v,
\label{EQ16a}\end{equation}
\begin{equation}
i\partial_T v = \left(-\nabla^2_{\vec{\rho}} +
\vec{\rho}\,{}^2\right)v + (\varepsilon +  a|v|^2)v - \kappa u.
\label{EQ16b}\end{equation}
\end{subequations}
Here
\begin{equation}
a =
\frac{a^{(v)}_s}{|a^{(u)}_s|}\frac{\int\mathrm{d}z|\psi_v|^4}
{\int\mathrm{d}z|\psi_u|^4},
\quad \kappa = \frac{2K}{\hbar\omega_\perp} =
\frac{2\varkappa}{1+\varkappa^2}\frac{E_1 -
E_0}{\hbar\omega_\perp},\quad \varepsilon =
\frac{2\mathcal{E}}{\hbar\omega_\perp} =
2\frac{1-\varkappa^2}{1+\varkappa^2}\frac{E_1 -
E_0}{\hbar\omega_\perp}\quad
 \label{EQ17}\end{equation}
The number of atoms $\mathcal{N}$ in the condensate  (the
$l_2$-norm) is given as follows
\begin{equation}
\mathcal{N}_{u,v} =  \int\mathrm{d}^2\vec{\rho}|\Phi_{u,v}|^2 =
\Delta N_{u,v},\quad N_u\equiv
\int\mathrm{d}^2\vec{\rho}|u|^2,\quad N_v\equiv
\int\mathrm{d}^2\vec{\rho}|v|^2.
 \label{EQ18}\end{equation}
The quantity $N_{u,v}$  will be referred to as the ``number of
atoms'' for short, since we are interested only in the relative
shares of the number of atoms in the two condensates and the ratio
of the total number of atoms to the collapse threshold in a single
2D NLS. The transformation coefficient $\Delta$  can be estimated
as $\frac{\ell_z}{8\pi|a_s|}$, it is of order $10^2-10^3$ for the
current trap sizes in the experiments with BECs.

The tunnelling coefficient $\kappa$ and the  zero-point energy
difference $\varepsilon$ of the dimensionless coupled-mode system
can take arbitrary values. Indeed, from the definition
(\ref{EQ17}) we have
\[
\kappa = \frac{2K}{\hbar\omega_\perp} =
\gamma\frac{2K}{\hbar\omega_z}, \quad \varepsilon =
\frac{2\mathcal{E}}{\hbar\omega_\perp} =
\gamma\frac{2\mathcal{E}}{\hbar\omega_z},
\]
i.e. there are two unrelated multipliers, the first, $\gamma$, is
large and the second is small.

Finally, we can reformulate the applicability condition
(\ref{EQ12}) for the coupled-mode system in the dimensionless
variables:
\begin{equation}
N_u \ll \gamma r^2_u,\quad aN_v \ll \gamma r^2_v,
 \label{EQ19}\end{equation}
where $r_{u,v} = d^{(u,v)}_\perp/\ell_\perp$ is the dimensionless
radius of the condensate. The condition (\ref{EQ19}) is derived by
using the transformation (\ref{EQ18}) with the estimate
$\int\mathrm{d}z|\psi_{u,v}|^4 \sim 1/\ell_z$ and that $\gamma
=\ell^2_\perp/\ell^2_z$.

We have verified that condition (\ref{EQ19}) is  satisfied (for
the pancake trap with $\gamma \ge 100$) by the stable stationary
states (figures 3,4,6,7 of section \ref{sec7}).

If more than two energy levels of the double-well trap are
significantly occupied by the condensate, the conditions for
applicability of the coupled-mode system (\ref{EQ13}) are
violated. In this case, one can use the nonlinear coupled-mode
approach \cite{Nonlmode} which results, however, in the
nonlinearly coupled NLS equations.

\section{Stability of the  stationary states}
\label{sec3}

BEC in a double-well trap can be unstable with respect to collapse
if the atomic interaction is attractive (the $u$-condensate in the
notations of the previous section). By setting $\kappa = 0$ in
system (\ref{EQ16}) we obtain for the $u$-condensate the focusing
2D NLS equation with an external potential. In the simplest case
when the potential is parabolic the collapse threshold is given by
the $l_2$-norm of the Townes soliton:
\begin{equation}
N_{\mathrm{th}} =  11.69,
  \label{EQ20}\end{equation}
since the collapse threshold is independent of the parabolic
potential \cite{LENS,Collapse2D}. The Townes soliton is the
solution $u = e^{iT}{R}(\rho)$ of the 2D NLS equation, i.e. the
function ${R}(\rho)$  satisfies
\begin{equation}
\nabla^2{R} + {R}^3 - {R} = 0.
 \label{EQ21}\end{equation}
Here and below the operator $\nabla$ is the gradient with respect
to $\vec{\,\rho}\,$ if is it not explicitly indicated otherwise by
a subscript. In the following we will need the well-known identity
for the Townes soliton (see, for instance, \cite{Collapse2D})
\begin{equation}
\int\mathrm{d}^2\vec{\rho}\,{R}^2 =
\int\mathrm{d}^2\vec{\rho}\,(\nabla{R})^2 =
\frac{1}{2}\int\mathrm{d}^2\vec{\rho}\,{R}^4.
  \label{EQ26}\end{equation}

Let us now consider the problem of stability of the stationary
states in the coupled-mode system. We are interested only in the
axially symmetric stationary states, $u = e^{-i\mu T}U(\rho)$ and
$v = e^{-i\mu T}V(\rho)$, where $\rho = |\vec{\rho}|$. The
stability or instability can be established by considering the
eigenvalue problem associated with the linearized system. Writing
the perturbed solution as follows
\begin{equation}
u = e^{-i\mu T}\left\{U(\rho) + e^{-i\Omega
T}\mathcal{U}(\vec{\rho})\right\},\quad v = e^{-i\mu
T}\left\{V(\rho) + e^{-i\Omega T}\mathcal{V}(\vec{\rho})\right\},
 \label{EQ27}\end{equation}
where $(\mathcal{U}$,$\mathcal{V})$  is a small perturbation mode
with the frequency $\Omega$, one arrives at the following linear
problem for the eigenfrequency:
\begin{equation}
-i\Omega\left(\begin{array}{c}\mathcal{U}_R\\\mathcal{V}_R\end{array}\right)
=\Lambda_0\left(\begin{array}{c}\mathcal{U}_I\\\mathcal{V}_I\end{array}\right),\quad
i\Omega\left(\begin{array}{c}\mathcal{U}_I\\\mathcal{V}_I\end{array}\right)
=\Lambda_1\left(\begin{array}{c}\mathcal{U}_R\\\mathcal{V}_R\end{array}\right),
 \label{EQ28}\end{equation}
with
\begin{equation}
\Lambda_0 = \left(\begin{array}{cc} L^{(u)}_0 & -\kappa \\
-\kappa & L^{(v)}_0\end{array}\right), \quad
\Lambda_1 = \left(\begin{array}{cc} L^{(u)}_1 & -\kappa \\
-\kappa & L^{(v)}_1\end{array}\right).
 \label{EQ29}\end{equation}
Here the scalar operators are defined as follows
\[
L^{(u)}_0 = -(\nabla^2 + U^2 + \mu) + \rho^2,\quad L^{(u)}_1 =
-(\nabla^2 + 3U^2 + \mu) + \rho^2,\quad L^{(v)}_0 = -(\nabla^2 -
aV^2 + \mu - \varepsilon) + \rho^2,
\]
\begin{equation}
\quad L^{(v)}_1 = -(\nabla^2  - 3aV^2 + \mu - \varepsilon) +
\rho^2.
 \label{EQ30}\end{equation}

First of all, the matrix operator $\Lambda_0$ is non-negative for
positive stationary solutions, i.e. satisfying $UV>0$. Indeed, the
scalar operators on the main diagonal of $\Lambda_0$ can be cast
as follows
\[
L^{(u)}_0 = -\frac{1}{U}\nabla U^2\nabla\frac{1}{U} +
\kappa\frac{V}{U},\quad L^{(v)}_0 = -\frac{1}{V}\nabla
V^2\nabla\frac{1}{V} + \kappa\frac{U}{V},
\]
what can be easily verified by direct calculation. Therefore the
scalar product of $\Lambda_0$ with any vector $X =
(X_1(\vec{\rho}),X_2(\vec{\rho}))$  is non-negative:
\[
\langle X|\Lambda_0|X\rangle = \int\mathrm{d}^2\vec{\rho}\,
\left\{X_1^*L^{(u)}_0X_1 + X_2^*L^{(v)}_0X_2 -
\kappa(X_1^*X_2+X_2^*X_1)\right\}
\]
\[
\ge \kappa\int\mathrm{d}^2\vec{\rho}\,\left|X_1\sqrt{\frac{V}{U}}
- X_2\sqrt{\frac{U}{V}}\,\right|^2 \ge 0.
\]
Here we have used the positivity  of the operators $
-\frac{1}{U}\nabla U^2\nabla\frac{1}{U} $ and $ -\frac{1}{V}\nabla
V^2\nabla\frac{1}{V}$. The operator $\Lambda_0$ has one zero mode
given by the stationary point itself: $Z = (U,V)$, $\Lambda_0Z=0$.

Non-negativity of $\Lambda_0$ is an essential property for the
following, therefore we will concentrate on the solutions
satisfying $UV>0$, which can be termed ``positive'' solutions,
while the ones, satisfying  $UV<0$, are discarded from the
consideration below. The latter solutions bifurcate from zero at a
higher energy then the positive ones (see  sections \ref{sec4} and
\ref{sec5}).

For a positive solution, the lowest eigenfrequency of the linear
stability problem can be found by minimizing the  quotient
\begin{equation}
\Omega^2 = \mathrm{min}\frac{\langle X|\Lambda_1|X\rangle}{\langle
X|\Lambda^{-1}_0|X\rangle}
 \label{EQ31}\end{equation}
in the space orthogonal to the zero mode of $\Lambda_0$: $\langle
Z|X\rangle = 0$ (here $\langle X|Y\rangle \equiv
\int\mathrm{d}^2\vec{\rho}\,(X^*_1Y_1 + Y_2X^*_2)$). Equation
(\ref{EQ31}) follows from the eigenvalue problem rewritten as
$\Lambda_0\Lambda_1X = \Omega^2X$ with $X =
(\mathcal{U}_R,\mathcal{V}_R)$.

The imaginary eigenfrequencies $\Omega$, which mean instability,
appear  due to negative eigenvalues of the operator $\Lambda_1$.
For the coupled-mode system without the transverse potential there
is at least one zero eigenvalue of $\Lambda_1$ due to the
translational invariance, with the eigenfunction (in the vector
form) being given as $Z_1 = (\nabla U,\nabla V)$. For the positive
stationary solutions, if there is only one negative eigenvalue,
the VK stability criterion $\frac{\partial N}{\partial \mu}<0$
applies, which can be established by a simple repetition of the
arguments presented, for instance,  in Ref. \cite{STAB}. The limit
on the number of negative eigenvalues is related to the fact that
the minimization of the quotient in equation (\ref{EQ31}) is
subject to only one orthogonality condition, thus only one
negative direction in the energy functional can be eliminated by
satisfying this condition.

The following simple strategy has been used to establish the
stability. The eigenvalue problem was reformulated in the polar
coordinates $(\rho,\theta)$ and the operators were expanded in
Fourier series with respect to $\theta$, which is done by the
simple substitution: $\nabla^2 \to \nabla^2_\rho - n^2/\rho^2$,
where $\nabla^2_\rho =
\partial^2_\rho +\rho^{-1}\partial_\rho$. Noticing that the
orbital operators $\Lambda_{1,n} = \Lambda_1(\nabla^2 \to
\nabla^2_\rho - n^2/\rho^2)$ are ordered as follows
$\Lambda_{1,n+1} \ge \Lambda_{1,n}$, we have checked for the
negative eigenvalues of  the first two orbital operators with $n =
0,1$. If there are two or more negative eigenvalues (for instance,
if $\Lambda_{1,0}$ and $\Lambda_{1,1}$ both have one negative
eigenvalue) then the solution is unstable; if there is only one
such of $\Lambda_{1,0}$ then one can apply the VK criterion. From
our numerical simulations it follows that $\Lambda_{1,1}$ is
always positive, while  the operator $\Lambda_{1,0}$ has one
negative eigenvalue or none (the latter case corresponds to the
defocusing effective nonlinearity  in the coupled-mode system with
the transverse potential, see sections \ref{sec5} and \ref{sec7}).

When the operator $\Lambda_1$ does not have negative eigenvalues
at all the solution is  unconditionally stable. In this case, the
formal threshold $\frac{\partial N}{\partial \mu} = 0$ of the
stability criterion  is the point where an additional zero mode of
the operator $\Lambda_{1,0}$ appears. Indeed, at the threshold
point we have
\[
\int\mathrm{d}^2\vec{\rho}\, \left(\frac{\partial U}{\partial
\mu}, \frac{\partial V}{\partial \mu}\right)\Lambda_{1,0}
\left(\begin{array}{c}\frac{\partial U}{\partial \mu}\\
\frac{\partial V}{\partial \mu}\end{array}\right) = 0,
\]
due to the identity
\begin{equation}
\Lambda_{1,0}\left(\begin{array}{c}\frac{\partial U}{\partial \mu}\\
\frac{\partial V}{\partial \mu}\end{array}\right) =
\left(\begin{array}{c} U\\ V\end{array}\right),
 \label{EQ32}\end{equation}
which follows from differentiation of the stationary coupled-mode
system, $\Lambda_0(U,V)^T = 0$, with respect to $\mu$. Hence, in
this case,  on one side of the VK threshold the operator
$\Lambda_{1,0}$ is positive and the solution is unconditionally
stable, while on the other side there is one negative eigenvalue
and the VK criterion applies (this corresponds  to a change of the
effective interaction from the repulsive to attractive, see
sections \ref{sec5} and \ref{sec7}).

Note that the above approach allows us to decide on the stability
of the stationary solutions to the coupled-mode system without
numerical solution of the full eigenvalue problem (\ref{EQ28}).

\section{Small-amplitude  and  asymptotic
solutions}
\label{bigsec4}

In this section we study two limiting cases of  solutions of the
coupled-mode system (\ref{EQ16}): the solution with vanishing
amplitude (i.e., the  bifurcation from zero) and the asymptotic
solution with the chemical potential taking large negative values.
There is an essential difference in the bifurcating
small-amplitude solutions in the coupled-mode systems with and
without the transverse potential, while the asymptotic solution,
though also possessing some minor difference between the two
cases, can be studied for both cases simultaneously. Accordingly,
the section is divided into three subsections. First we study the
soliton bifurcation from zero, i.e. the coupled-mode system
without the transverse potential, in subsection \ref{sec4}. In
subsection \ref{sec5}, the same bifurcation is considered in the
coupled-mode system with a general transverse potential. We derive
the asymptotic solution in subsection \ref{sec6}  using, for
simplicity, the parabolic transverse potential.

\subsection{The soliton bifurcation from zero }
\label{sec4}

Though the linear coupling of the repulsive and attractive  NLS
equations  breaks the scale-invariance, the  2D-soliton solutions
are always unstable. This conclusion follows from a comparison of
the soliton bifurcation from zero, i.e. $\mu\to \mu_\mathrm{bif}$
(which is the subject of this section) and the asymptotic solution
when $\mu \to -\infty$ (which is considered in section
\ref{sec6}). Moreover it is supported by the direct numerical
solution of section \ref{sec7}.

The stationary coupled-mode system (when the transverse potential
is flat) reads
\begin{subequations}
\label{EQ33}
\begin{eqnarray}
 \mu U + \nabla^2 U + U^3 +\kappa V &=& 0,\label{EQ33a}\\
(\mu - \varepsilon) V + \nabla^2 V - aV^3 +\kappa U &=& 0.
\label{EQ33b}
\end{eqnarray}
 \end{subequations}
First of all, the most important information on the existence of
solitons is provided by the dispersion law of the linearized
system, which has two branches in the case of the coupled-mode
system (\ref{EQ33}). Setting $U = U_0e^{-\lambda \rho}$ and $V =
V_0e^{-\lambda\rho}$, in the limit of vanishing $U_0$ and $V_0$ we
obtain:
\begin{equation}
\lambda^2_{1,2} = \mu_{1,2} - \mu,\quad \mu_{1} =
\frac{\varepsilon}{2} - \sqrt{\frac{\varepsilon^2}{4} +
\kappa^2},\quad \mu_{2} = \frac{\varepsilon}{2} +
\sqrt{\frac{\varepsilon^2}{4} + \kappa^2}.
 \label{EQ34}\end{equation}
It is easy to see that the following inequalities hold: $\mu_1 < 0
< \mu_2$ and $\mu_1<\varepsilon < \mu_2$.

The same dispersion relations appear also in the 1D case as well
\cite{PRA}, their universality is due to the fact that the
limiting values of the chemical potential are determined by the
energy difference in the double-well trap and the trap asymmetry:
\begin{equation}
\mu_1=\left(\frac{1-\varkappa^2}{1+\varkappa^2}-\frac{1}{2}\right)
\frac{E_1-E_0}{\hbar\omega_\perp},\quad
\mu_2=\left(\frac{1-\varkappa^2}{1+\varkappa^2}+\frac{1}{2}\right)
\frac{E_1-E_0}{\hbar\omega_\perp}.
 \label{EQ35}\end{equation}

In the 1D case the positive solitons ($UV>0$) bifurcate from zero
at the lower energy level $\mu = \mu_1$, while the non-positve
ones ($UV<0$) from the higher level $\mu= \mu_2$. This property
also holds in the 2D case. This can be shown as follows. First of
all, only one branch of soliton solutions may correspond to each
branch of the dispersion law. Second, there is a point $\rho_0$ on
the positive real line such that $\nabla^2 U(\rho_0) = 0$. Indeed,
setting $\xi = \ln\rho$ we have $\nabla^2 U(\rho) =
e^{-2\xi}\frac{d^2 U}{d\xi^2}$, whereas, for the solution with a
finite $l_2$-norm, the first derivative $\frac{d U}{d\xi} = \rho
\frac{d U}{d\rho}$ tends to zero as $\xi\to\pm\infty$  (i.e., when
$\rho\to0$ or $\rho\to \infty$). Considering  equation
(\ref{EQ33a}) at $\rho =\rho_0$ we obtain $(\mu U + U^3 + \kappa
V)|_{\rho_0} = 0$, thus $\mu<0$ for $UV>0$ (recall that $\kappa
>0$), i.e. the positive solitons belong to the $\lambda_1$-branch of the
dispersion law, while the non-positive ones to the
$\lambda_2$-branch.

Consider now the positive solitons with vanishing amplitude, i.e.
in the limit $\mu\to \mu_{\mathrm{bif}} \equiv \mu_1$. Set $\mu =
\mu_1 - \epsilon$, with $\epsilon\to 0$ and suppose that in this
limit $U = \mathcal{O}(\epsilon^p)$ and $V =
\mathcal{O}(\epsilon^q)$ with some $p,q>0$. From  system
(\ref{EQ33}) we necessarily get $p = q$. Further steps are
essentially the same as in the 1D case \cite{PRA}. First we expand
$V$ in  the power series with respect to $U$ and its derivatives,
using for this goal equation (\ref{EQ33b}):
\begin{equation}
V =\kappa\beta U - a\kappa^3\beta^4 U^3 +3a^2\kappa^5\beta^7U^5
+\kappa\beta^2\nabla^2 U + \mathcal{O}(\epsilon^s),
 \label{EQ36}\end{equation}
where $\beta = (\varepsilon - \mu)^{-1}$ and  $s =
\mathrm{min}\{3p+1,p+2,7p\}$. Here  $\beta$ should also be
expanded with respect to $\epsilon$. Second, the result is
substituted into  equation (\ref{EQ33a}) and we get
\begin{equation}
-\epsilon\left(1 + \frac{\kappa^2}{\mu_2^2}\right)U + \left(1 +
\frac{\kappa^2}{\mu_2^2}\right)\nabla^2U + \left(1 -
\frac{a\kappa^4}{\mu_2^4}\right)U^3 +
\frac{3a^2\kappa^6}{\mu_2^7}U^5 = \mathcal{O}(\epsilon^s),
 \label{EQ37}\end{equation}
From the coefficient at the  cubic term and the definition of
$\mu_2$ one immediately concludes that the soliton bifurcation
from zero is possible if and only if $\varepsilon \ge
\varepsilon_\mathrm{cr}$, where $\varepsilon_\mathrm{cr}$ is the
same as in the 1D case \cite{PRA}:
\begin{equation}
\varepsilon_\mathrm{cr} = \kappa\left(a^{1/4} - a^{-1/4}\right).
 \label{EQ38}\end{equation}

For $\varepsilon > \varepsilon_\mathrm{cr}$ one can drop the
quintic term from equation (\ref{EQ37}), thus the effective
bifurcation equation  is the canonical 2D NLS equation except for
the coefficients. In this case the condition that all terms have
the same order requires that $p = 1/2$ and $\nabla^2\sim
\epsilon$, i.e. there is a new length scale $\xi =
\sqrt{\epsilon}\rho$. The soliton solution reads
\begin{equation}
U = \sqrt{\epsilon}A_0\left\{{R}(\sqrt{\epsilon}\rho) + \epsilon
\mathcal{G}(\sqrt{\epsilon}\rho) +
\mathcal{O}(\epsilon^2)\right\},\quad A_0 \equiv \left(1 +
\frac{\kappa^2}{\mu_2^2}\right)^{1/2}\left(1 -
\frac{a\kappa^4}{\mu_2^4}\right)^{-1/2}.
 \label{EQ39}\end{equation}
Here ${R}(\xi)$  is the Townes soliton.

While the soliton amplitude approaches zero, the corresponding
number of atoms has a \textit{finite} limit. Indeed, we have
\begin{equation}
N_u = \int\mathrm{d}^2\vec{\rho}\, U^2 = A_0^2N_\mathrm{th} +
\mathcal{O}(\epsilon),
 \label{EQ40}\end{equation}
with $N_\mathrm{th}$ being the threshold for collapse (\ref{EQ20})
in a single 2D NLS equation. Thus, the bifurcation from zero is
discontinuous in the 2D case. This is quite dissimilar to the 1D
case, where only on the boundary $\varepsilon =
\varepsilon_\mathrm{cr}$ the soliton bifurcation from zero  is
discontinuous \cite{PRA}.

On the boundary $\varepsilon = \varepsilon_\mathrm{cr}$ the
effective equation is the quintic NLS equation and $p = 1/4$,
exactly as in the 1D case. However, in contrast to the 1D case,
the bifurcation from zero is singular. Now $U =
\epsilon^{1/4}B_0(F_0(\sqrt{\epsilon}\rho) + \epsilon
F_1(\sqrt{\epsilon}\rho) + \mathcal{O}(\epsilon^2))$, with some
$B_0$ and $F_0(\xi)$. Hence, the number of atoms $N_u \propto
\epsilon^{-1/2}$ as $\epsilon \to 0$. The bifurcating solitons are
 unstable in this case by the VK criterion.

For $\varepsilon > \varepsilon_\mathrm{cr}$, to decide on the
stability of the soliton solutions  near the bifurcation point one
has to find the derivative of the number of atoms with respect to
$\epsilon$ in the limit $\epsilon\to 0$. To this end the soliton
solution up to the order $\epsilon^{3/2}$ is required. The
necessary higher-order expansion for $V$ in terms of  $U$   reads
\begin{equation}
V =\kappa\beta U - a\kappa^3\beta^4 U^3 +3a^2\kappa^5\beta^7U^5
+\kappa\beta^2\nabla^2 U +\kappa\beta^3\nabla^4U
-2a\kappa^3\beta^5\nabla^2U^3 + \mathcal{O}(\epsilon^{7/2}).
 \label{EQ41}\end{equation}
Substituting  this expression into equation (\ref{EQ33a}) one can
get an equation for the  correction $\mathcal{G}=\mathcal{G}(\xi)$
to the soliton solution (\ref{EQ39}). The result is
\begin{equation}
\mathcal{L}_1\mathcal{G} \equiv (1 - \nabla^2_{\vec{\xi}} -
3{R}^2)\mathcal{G} = \mathcal{L}_2{R},
 \label{EQ42}\end{equation}
with the operator $\mathcal{L}_2$ defined as follows
\begin{equation}
\mathcal{L}_2 \equiv
\left(1+\frac{\kappa^2}{\mu_2^2}\right)^{-1}\frac{\kappa^2}{\mu_2^3}\left\{
1-2\left(1+\frac{a\kappa^2}{\mu_2^2}A_0^2\right)\nabla_{\vec{\xi}}^2
+\left(1+2\frac{a\kappa^2}{\mu_2^2}A_0^2\right)\nabla_{\vec{\xi}}^4\right\}.
 \label{EQ43}\end{equation}
Now, we can find the total number of atoms $N = N_u + N_v$. We
have
\begin{equation}
N_u = A_0^2N_\mathrm{th} + 2\epsilon
A_0^2\int\mathrm{d}^2\vec{\xi}\,{R}\mathcal{G} +
\mathcal{O}(\epsilon^2).
 \label{EQ44}\end{equation}
The $v$-component of the soliton up to the order $\epsilon^{3/2}$
follows from  equations (\ref{EQ39}) and (\ref{EQ41}):
\begin{equation}
V = \sqrt{\epsilon}A_0\frac{\kappa}{\mu_2}\left\{ {R}
  + \epsilon\left(
\frac{1}{\mu_2}(\nabla_{\vec{\xi}}^2 - 1){R} -
A_0^2\frac{a\kappa^2}{\mu_2^3}{R}^3 + \mathcal{G} \right) +
\mathcal{O}(\epsilon^{2})\right\}.
 \label{EQ45}\end{equation}
Thus the number of atoms $N_v$ is given as
\begin{equation}
N_v = A_0^2\frac{\kappa^2}{\mu_2^2}N_\mathrm{th} +
2\epsilon\frac{\kappa^2}{\mu_2^2}A_0^2\mathcal{I} +
\mathcal{O}(\epsilon^{2}),
 \label{EQ46}\end{equation}
 where we have denoted
\[
\mathcal{I} \equiv \int\mathrm{d}^2\vec{\xi}\,{R}\left(
\frac{1}{\mu_2}(\nabla_{\vec{\xi}}^2 - 1){R} -
A_0^2\frac{a\kappa^2}{\mu_2^3}{R}^3 + \mathcal{G} \right) =
\int\mathrm{d}^2\vec{\xi}\,\left\{{R}\mathcal{G}
-\frac{2}{\mu_2}\left(1 +
\frac{a\kappa^2}{\mu_2^2}A_0^2\right){R}^2\right\}.
\]
Relation (\ref{EQ26}) for the Townes soliton has been used to
simplify the above expression for $\mathcal{I}$.

The scalar product of ${R}$ and $\mathcal{G}$, which enters the
expression for the number of atoms, is, in fact, equal to zero.
Indeed, using equation (\ref{EQ42}) we get
\[
\int\mathrm{d}^2\vec{\xi}\,{R}\mathcal{G} =
\int\mathrm{d}^2\vec{\xi}\,{R}\mathcal{L}_1^{-1}\mathcal{L}_2{R},
\]
but
\begin{equation}
\mathcal{L}_1^{-1}{R} = -\frac{1}{2}\left({R} +
\vec{\xi}\nabla_{\vec{\xi}}R\right).
 \label{EQ47}\end{equation}
(Equation can be obtained by taking the derivative of the
stationary 2D NLS equation with respect to chemical potential
$\mu$ at $\mu=-1$.) Therefore
\[
\int\mathrm{d}^2\vec{\xi}\,{R}\mathcal{G} =
-\frac{1}{2}\int\mathrm{d}^2\vec{\xi}\,\left({R}\mathcal{L}_2{R}
+\frac{1}{2}\vec{\xi}\nabla_{\vec{\xi}}(R\mathcal{L}_2 R)\right) =
0
\]
via integration by parts in the second term. Hence,  the total
number of atoms assumes the following form
\begin{equation}
N = N_u + N_v =  \left(1 +
\frac{\kappa^2}{\mu_2^2}\right)A_0^2N_\mathrm{th} - 4\epsilon
A_0^2 \frac{\kappa^2}{\mu_2^3}\left(1 +
\frac{a\kappa^2}{\mu_2^2}A_0^2\right)N_\mathrm{th} +
\mathcal{O}(\epsilon^2).
 \label{EQ48}\end{equation}
Clearly, (for $a\ge0$) in the vicinity of the bifurcation point
$\mu=\mu_1$ we have  $\frac{\partial N}{\partial \mu} =
-\frac{\partial N}{\partial \epsilon}>0$ which renders the
two-dimensional bifurcating solitons unstable  in contrast to
the stability of the similar bifurcating solitons in one spatial
dimension \cite{PRA}.

\subsection{Bifurcation from zero in the presence of transverse potential}
\label{sec5}

We have seen  that the soliton bifurcation from zero is always
discontinuous due to the fact that the solution has a new length
scale $\xi=\sqrt{\epsilon}\rho$. If the transverse parabolic
potential is taken into account (i.e., when it is not flat) the
bifurcating solution has the length scale of order 1 (i.e., the
order of the oscillator length).  The stationary coupled-mode
system with the parabolic transverse potential can be cast in a
form analogous to that of  system (\ref{EQ33}):
\begin{subequations}
\label{EQ49}
\begin{eqnarray}
 \omega U + \mathcal{D} U + U^3 +\kappa V &=& 0,\label{EQ49a}\\
(\omega - \varepsilon) V + \mathcal{D} V - aV^3 +\kappa U &=& 0,
\label{EQ49b}
\end{eqnarray}
 \end{subequations}
where we have introduced the operator $\mathcal{D}= \nabla^2 + 2 -
\rho^2$ and a shifted chemical potential $\omega = \mu -2$. Note
that $\mathcal{D}\le0$ with $\mathcal{D}e^{-\rho^2/2} = 0$. System
(\ref{EQ49}) in the limit of a vanishing amplitude solution, $U =
Ae^{-\rho^2/2}$ and $V = Be^{-\rho^2/2}$, with $A,B\to0$, gives
two boundary values of $\omega$ which coincide with the limiting
chemical potentials of section \ref{sec4}: $\omega_{1,2} =
\mu_{1,2}$. Moreover, the amplitudes are related as follows
$B_{1,2} = - (\omega_{1,2}/\kappa)A_{1,2}$. Therefore, the
positive solution $AB>0$ bifurcates at $\omega = \omega_1$,
exactly as in the soliton case.

To study the bifurcation in detail let us set $\omega = \omega_1 -
\epsilon$ with $\epsilon \to 0$ (here $\epsilon$ can be negative).
Further steps in the derivation of the leading order equation for
the bifurcating solution are formally the same as those in section
\ref{sec4}, one has only to substitute $\mu\to \omega$ and
$\nabla^2\to\mathcal{D}$.  For instance, we have   $U \sim V \sim
|\epsilon|^{p}$, where $p>0$. For $V$ we obtain the expression
formally equivalent to that of equation (\ref{EQ36}),
\begin{equation}
V =\kappa\beta U - a\kappa^3\beta^4 U^3 +\kappa\beta^2\mathcal{D}U
+ 3a^2\kappa^5\beta^7U^5 + \mathcal{O}(|\epsilon|^{s}),
 \label{EQ50}\end{equation}
with $\beta = (\varepsilon - \omega)^{-1}$ and $s =
\mathrm{min}\{3p+1,p+2,7p\}$,  while $U$ satisfies the equation
\begin{equation}
-\epsilon\left(1 + \frac{\kappa^2}{\mu_2^2}\right)U + \left(1 +
\frac{\kappa^2}{\mu_2^2}\right)\mathcal{D}U + \left(1 -
\frac{a\kappa^4}{\mu_2^4}\right)U^3 +
\frac{3a^2\kappa^6}{\mu_2^7}U^5  = \mathcal{O}(|\epsilon|^{s}).
 \label{EQ51}\end{equation}

In the derivation of  equations (\ref{EQ50}) and (\ref{EQ51}) it
is assumed that $\mathcal{D}\sim \epsilon$, an analog of
$\nabla^2\sim\epsilon$ of the previous section, though the reason
is different: the operator $\mathcal{D}$ is small because of its
discrete spectrum and the fact that the solution  bifurcates from
the ground state (with zero eigenvalue). The critical zero-point
energy difference $\varepsilon_\mathrm{cr}$ (\ref{EQ38})
delineates the regions of the defocusing and focusing cases;
though for $\varepsilon<\varepsilon_{\mathrm{cr}}$ the effective
equation (\ref{EQ51}) is defocusing, it has a localized solution
thanks to the external potential (in the operator $\mathcal{D}$).

To make our approach general we will use only two properties of
the operator $\mathcal{D}$, namely that it is non-positive (with
zero being an eigenvalue) and that it has a discrete spectrum. The
case of $\varepsilon = \varepsilon_\mathrm{cr}$ is a special case
of the bifurcation from zero, exactly as for the soliton
bifurcation of the previous section. Consider first $\varepsilon
\ne \varepsilon_\mathrm{cr}$, i.e. when the cubic term in equation
(\ref{EQ51}) has a non-zero coefficient (positive or negative).
Hence,  $p = 1/2$ in this case. Defining a new dependent variable
$F$ by setting $U = \sqrt{|\epsilon|}|A_0|F(\rho)$, with $A_0$
given by equation (\ref{EQ39}),  we obtain for $F$ the equation
\begin{equation}
- F + \epsilon^{-1}\mathcal{D}F + \sigma F^3   =
\mathcal{O}(|\epsilon|^{3/2}),\quad \sigma =
\mathrm{sgn}\left\{\epsilon(\varepsilon -
\varepsilon_{\mathrm{cr}})\right\}, \label{EQ52}\end{equation}
which has  a non-zero solution in the leading order $\epsilon^0$
only if $\sigma>0$, i.e. when the sign of $\epsilon$ is equal to
that of $\varepsilon-\varepsilon_\mathrm{cr}$. Equation
(\ref{EQ52}) allows one to obtain two leading orders of $F$ in the
expansion with respect to $\epsilon$: $F = F_0 + \epsilon F_1 +
\mathcal{O}(\epsilon^2)$. The simplest way to get them is to
invert the operator $1-\epsilon^{-1}\mathcal{D}$:
\begin{equation}
(1-\epsilon^{-1}\mathcal{D})^{-1} = |\phi_0\rangle\langle\phi_0| +
\epsilon\sum_{n=1}^\infty\frac{|\phi_n\rangle\langle\phi_n|}{\lambda_n
+\epsilon}\, \label{EQ53}\end{equation} where we have used the
eigenvalues, $-\lambda_n$ ($\lambda_n>0$ for $n\ge1$), and the
eigenfunctions, $|\phi_n\rangle$, of the operator $\mathcal{D}$:
$\mathcal{D}|\phi_n\rangle = -\lambda_n|\phi_n\rangle$. Now,
multiplying  equation (\ref{EQ52}) by the operator from equation
(\ref{EQ53}) and collecting the successive orders of $\epsilon$ we
obtain
\begin{equation}
F = \langle\phi_0^4\rangle^{-1/2}\left\{1 -
\frac{3\epsilon}{2\langle\phi_0^4\rangle^{2}}
\sum_{n=1}^\infty\frac{\langle\phi^3_0\phi_n\rangle^2}{\lambda_n}\right\}
\phi_0(\rho) + \epsilon\langle\phi_0^4\rangle^{-3/2}
\sum_{n=1}^\infty\frac{\langle\phi^3_0\phi_n\rangle^2}{\lambda_n}\phi_n(\rho)
+ \mathcal{O}(\epsilon^{2}).
\label{EQ54}\end{equation}
Here $\langle \ldots \rangle$ denotes the integral
$\int\mathrm{d}^2\vec{\rho} (\ldots)$.

Let us find the number of atoms corresponding to the bifurcating
solution. We need just the leading order, since,
 the bifurcation is continuous and the number of atoms
tends to zero as $\omega\to \omega_1$. We get
\begin{equation}
N  = \int\mathrm{d}^2\vec{\rho}\, (U^2 + V^2) =
|\epsilon|\langle\phi_0^4\rangle^{-1}\left(1+\frac{\kappa^2}{\mu_2^2}\right)A_0^2
+\mathcal{O}(\epsilon^2), \label{EQ55}\end{equation} where we have
used that in the leading order $V = \frac{\kappa}{\mu_2}U +
\mathcal{O}(|\epsilon|^{3/2})$. Now, for $\epsilon>0$ we have
$\frac{\partial N}{\partial \mu} = - \frac{\partial N}{\partial
\epsilon} <0$, i.e., the solution is stable for
$\varepsilon>\varepsilon_\mathrm{cr}$ by the VK criterion (since
there is only one negative eigenvalue of the operator $\Lambda_1$
(\ref{EQ29})). In the
 case  $\varepsilon<\varepsilon_\mathrm{cr}$ the
solution is unconditionally stable (the operator $\Lambda_1$ is
positive). Unconditional stability  in the latter case can be
explained by the fact that the effective equation (\ref{EQ51}) is
the defocusing NLS equation with the external potential, hence the
ground state solution is unconditionally stable.

Consider now the special case of the bifurcation from zero, when
$\varepsilon=\varepsilon_\mathrm{cr}$. The leading order
nonlinearity is quintic in $U$ and we get $p = 1/4$. Defining a
new variable $F$  by setting $U = |\epsilon|^{1/4}B_0F(\rho)$,
where $B^4_0 =
\left(1+\frac{\kappa^2}{\mu_2^2}\right)\frac{\mu_2^7}{3a^2\kappa^6}$,
we get the effective equation
\begin{equation}
- F + \epsilon^{-1}\mathcal{D}F + \mathrm{sgn}(\epsilon)F^5   =
\mathcal{O}(|\epsilon|^{3/4})
\label{EQ56}\end{equation}
which allows one to obtain the leading order $F = F_0
+\mathcal{O}(\epsilon)$.  It is clear that in this case $\epsilon
> 0$.  We get $F_0 =
\langle\phi_0^6\rangle^{-1/4}\phi_0(\rho)$. Finally, using  $V
= \frac{\kappa}{\mu_2}U + \mathcal{O}(\epsilon^{3/4})$, we obtain
\begin{equation} N = N_u + N_v =
\sqrt{\epsilon}\langle\phi_0^6\rangle^{-1/2}
\left(1+\frac{\kappa^2}{\mu_2^2}\right)B^2_0
+\mathcal{O}(\epsilon).
\label{EQ57}\end{equation}
Therefore in this case the solution is also stable by the VK
criterion (there is only one negative eigenvalue of the operator
$\Lambda_{1}$ (\ref{EQ29}), similar as in the previous case).

We see that  the confining transverse potential  allows for stable
small-amplitude solutions. Moreover, for the zero-point energy
difference below the critical value,
$\varepsilon<\varepsilon_\mathrm{cr}$, the effective equation for
the bifurcating solution is  the defocusing NLS equation, whereas
in the opposite case it is the focusing cubic or quintic NLS
equation.

\subsection{Asymptotic solution of the coupled-mode system
for {\large $\mu\to-\infty$ }}
\label{sec6}

In the previous two sections we have considered the bifurcation
from zero. To complete the consideration, one has to study also
the other limit of the chemical potential, i.e. $\mu\to-\infty$.
This can be done in a unified way for the coupled-mode system with
or without the transverse potential. The reason is that in both
cases there is the same new length scale and the solution is
approximated by the Townes soliton.  Therefore, we will use the
stationary system (\ref{EQ49}) to study the  asymptotic solution.
Thus we set $\omega = -\epsilon^{-1}$ with $\epsilon\to 0$, where
$\omega = \mu-2$ as in the previous section. Finally, we will
restrict the consideration to the case  $a\ge0$, which is the most
interesting one, and assume that the transverse potential is
parabolic (which is not an essential requirement, but is
convenient for calculations).

Analysis of the system (\ref{EQ49}) reveals that in the limit
$\omega\to-\infty$ the leading order of the solution is as
follows: $U = \mathcal{O}(\epsilon^{-1/2})$ and $V =
\mathcal{O}(\epsilon^{1/2})$. Indeed,  the leading order of $V$
cannot be greater then that of $U$ otherwise there is no localized
solution  (for $a\ge0$) in the limit $\epsilon\to0$. Equation
(\ref{EQ49a}) for $U$ has a non-trivial solution only if
$\mathcal{D} = \mathcal{O}(\epsilon^{-1})$. The only way to
satisfy the latter is to require that $\nabla^2\sim
\epsilon^{-1}$, i.e. there is a new length scale
$\xi=\epsilon^{-1/2}\rho$ (the effective length of the solution
tends to zero as $\omega\to-\infty$ and the external potential is
negligible). We will need to compute $U$ up to the order
$\epsilon^{3/2}$ and $V$ up to the order $\epsilon^{1/2}$. Setting
\[
U = \epsilon^{-1/2}(U^{(0)}(\epsilon^{-1/2}\rho) + \epsilon
U^{(1)}(\epsilon^{-1/2}\rho) +  \epsilon^2
U^{(2)}(\epsilon^{-1/2}\rho) + \mathcal{O}(\epsilon^{3})),
\]
\[
V = \epsilon^{1/2}(V^{(0)}(\epsilon^{-1/2}\rho) +
\mathcal{O}(\epsilon)),
 \]
and expanding system (\ref{EQ49}) in the series with respect to
$\epsilon$, we obtain  $U^{(0)} = R(\xi)$  and
\begin{subequations}
 \begin{equation}
\mathcal{L}_1 U^{(1)}(\xi)  = 2R(\xi),
 \label{EQ58a}\end{equation}
\begin{equation}
 \mathcal{L}_1U^{(2)}(\xi) = 2U^{(1)}(\xi) - \xi^2R(\xi) +
3R(\xi)U^{(1)}(\xi) +\kappa V^{(0)}(\xi),
 \label{EQ58b}\end{equation}
\begin{equation}
(1 - \nabla^2_{\vec{\xi}})V^{(0)}(\xi)  = \kappa R(\xi),
 \label{EQ58c}\end{equation}
\end{subequations}
where the operator $\mathcal{L}_1$ is given in equation
(\ref{EQ42}). Since in the leading order we have the Townes
soliton, $\epsilon$ is positive, i.e. the solutions with
$\omega\to\infty$ are impossible (though for
$\varepsilon<\varepsilon_\mathrm{cr}$, according to the results of
section \ref{sec5},   the curve $N = N(\omega)$ corresponding the
solution of the coupled-mode system with the transverse potential
initially enters the right half-plane $\omega>\omega_1$, it
eventually turns left and approaches $-\infty$, see figure
\ref{FG5} in the next section).

By setting $\kappa = 0$ in equation (\ref{EQ58b}) one must obtain
the derivative $\frac{\partial N}{\partial \mu}$ corresponding to
a single NLS equation (with or without the transverse potential)
in the limit $\mu\to-\infty$. In the case without the transverse
potential we know that this derivative is zero due to the critical
scale invariance. In the case of a single 2D NLS equation with an
external potential the derivative is negative and the number of
atoms approaches the collapse threshold $N_\mathrm{th}$
(\ref{EQ20}) from below. ``Switching on'' the quantum tunnelling
changes this behavior: for $\kappa> \kappa_\mathrm{cr}$, with some
$\kappa_\mathrm{cr}$, the derivative $\frac{\partial N}{\partial
\mu}$ assumes  a positive value and the number of atoms approaches
the collapse threshold $N_\mathrm{th}$  from above. Let us find
the critical tunnelling coefficient. We have:
\[
N_u = \int\mathrm{d}^2\vec{\rho}\,U^2 = N_\mathrm{th} +
\epsilon^2\int\mathrm{d}^2\vec{\xi}\, \left\{{U^{(1)}}^2 +
2RU^{(2)}\right\} + \mathcal{O}(\epsilon^3)
\]
\[
= N_\mathrm{th} -
\epsilon^2\int\mathrm{d}^2\vec{\xi}\,\left\{\vec{\xi}{\,}^2R^2\right\}
+\epsilon^2\kappa^2\int\mathrm{d}^2\vec{\xi}\,U^{(1)}(1-\nabla^2_{\vec{\xi}})^{-1}R
+ \mathcal{O}(\epsilon^3), \]
\[
N_v = \int\mathrm{d}^2\vec{\rho}\,V^2 =
\epsilon^2\kappa^2\int\mathrm{d}^2\vec{\xi}\,R(1-\nabla^2_{\vec{\xi}})^{-2}R
+ \mathcal{O}(\epsilon^3).
\]
In the derivation of these formulae we have solved equation
(\ref{EQ58b}) for $U^{(2)}$ and (\ref{EQ58c}) for $V^{(0)}$ by the
inversion of the corresponding operators and integration by parts.
We have also used that the scalar product
$\int\mathrm{d}^2\vec{\xi}\,RU^{(1)}$ is equal to zero, which is
an immediate consequence of the fact that equation (\ref{EQ58a})
for $U^{(1)}$ does not involve the external potential  (equation
(\ref{EQ47}) also can be used to establish this fact directly).
Using equation (\ref{EQ58a}) and integrating by parts to get rid
of the term with $U^{(1)}$ in the last integral in the expression
for $N_u$, we get the following formula for the total number of
atoms
\begin{equation}
N = N_u + N_v =  N_\mathrm{th} -
\epsilon^2\int\mathrm{d}^2\vec{\xi}\, {\xi}^2 R^2
+\epsilon^2\kappa^2\int\mathrm{d}^2\vec{\xi}\,R(1-\nabla^2_{\vec{\xi}})^{-1}R
+ \mathcal{O}(\epsilon^3).
\label{EQ59}\end{equation}

For the coupled-mode system without the transverse potential, the
first integral on the r.h.s. of equation (\ref{EQ59})  is absent.
The second integral is positive:
\begin{equation}
I_1\equiv
\int\mathrm{d}^2\vec{\xi}\,R(1-\nabla^2_{\vec{\xi}})^{-1}R \approx
7.41.
\label{EQ60}\end{equation}
Hence, in this case, the number of atoms always approaches the
threshold for collapse $N_\mathrm{th}$ of a single 2D NLS equation
from above.

Consider now  the case of the parabolic transverse potential. In
this case the second term on the r.h.s. of equation (\ref{EQ59})
is negative and
\[
I_2\equiv \int\mathrm{d}^2\vec{\xi}\, {\xi}^2 R^2 \approx 13.82.
\]
Therefore, the two terms of order $\epsilon^2$  on the r.h.s. of
equation (\ref{EQ59}) compensate each other at some $\kappa =
\kappa_\mathrm{cr}$ and the  derivative $\frac{\partial
N}{\partial \mu}$ changes sign in the limit $\mu\to-\infty$. The
derivative itself is, however, of the order $-1/\omega^3 =
\epsilon^3$ and the change of sign is not visible numerically.

The important conclusion from the asymptotic solution of the
coupled-mode system for $\mu\to-\infty$ is that a new scale
appears and the $u$-component of the solution is approximated by
the Townes soliton, while the amplitude of the $v$-component tends
to zero. Hence, there are always solutions which suffer from the
collapse instability. They have the total number of atoms
approaching $N_\mathrm{th}$ as $\mu\to-\infty$. The collapse
instability also appears in this limit in the one-dimensional
coupled-mode system \cite{PRA}, where for a large number of atoms
the solution with $\mu\to-\infty$ is, in fact, the collapsing
ground state. However, in the two-dimensional system with the
transverse potential another stationary solution becomes the
ground state (see the next section).

\section{Numerical solution of the coupled-mode system.}
\label{sec7}

\begin{figure}[hbp]
\includegraphics{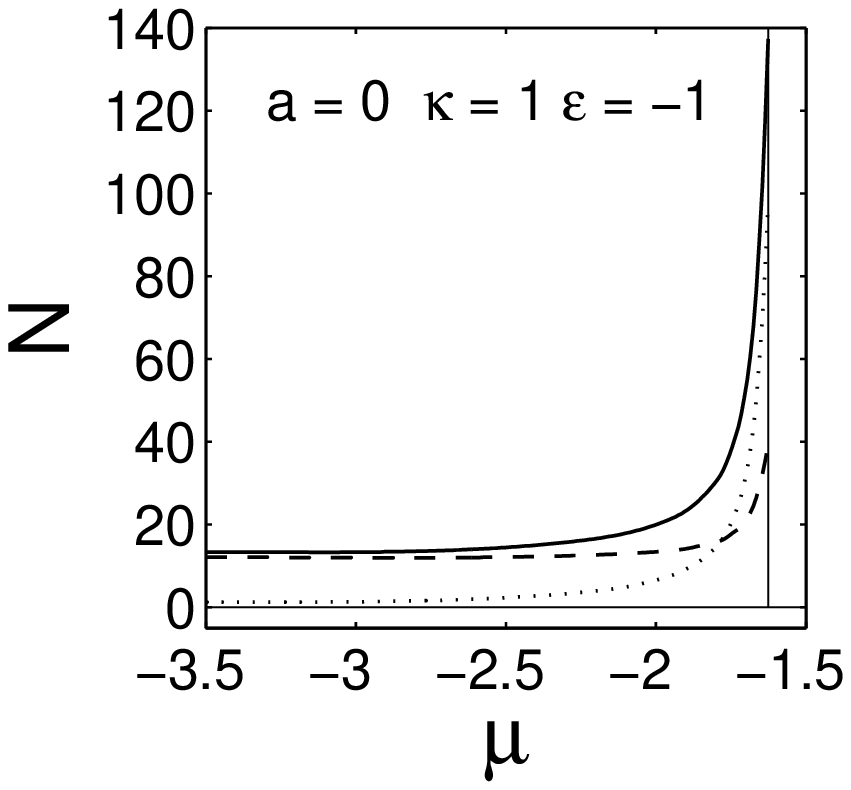}
\caption{\label{FG1} The total number of atoms (solid line)  vs.
the chemical potential corresponding to the 2D-soliton solutions
of the coupled-mode system.  The dashed and dotted lines give the
number of atoms in the $u$- and $v$-condensates, respectively. The
number of atoms here is reduced by the factor $\Delta$, as in
equation (\ref{EQ18}). The chemical potential is given in the
units of $\hbar\omega_\perp/2$. }
\end{figure}

From the analytical study of the bifurcation from zero and the
asymptotic solution for $\mu\to-\infty$ we can conclude the
following. First of all, in the case of the coupled-mode system
without the transverse potential, the  soliton solutions exist for
$\mu\le\mu_\mathrm{bif}$ with $\mu_\mathrm{bif}\equiv \mu_1$ from
equation (\ref{EQ34}). The
 expansions of the total number of atoms for the
bifurcating and asymptotic solitons are as follows:
\begin{equation}
N = \left(1 + \frac{\kappa^2}{\mu_2^2}\right)A_0^2N_\mathrm{th}+
4(\mu-\mu_\mathrm{bif}) A_0^2 \frac{\kappa^2}{\mu_2^3}\left(1 +
\frac{a\kappa^2}{\mu_2^2}A_0^2\right)N_\mathrm{th} + \mathcal{O}(
\mu-\mu_\mathrm{bif})^2,\quad \mu\to\mu_\mathrm{bif},
\label{EQ61}\end{equation}
\begin{equation}
N=N_\mathrm{th}+\frac{\kappa^2}{\mu^2}I_1 +
\mathcal{O}(\mu^{-3}),\quad \mu \to -\infty,
\label{EQ62}\end{equation}
where $A_0$  is defined in equation (\ref{EQ39}), $R(\xi)$ is the
Townes soliton and $I_1$ is given by formula (\ref{EQ60}). Hence,
we have $N(\mu_\mathrm{bif})
> N_\mathrm{th}$ for all values of the system parameters (we consider $a\ge0$).
Moreover, in both limits  the derivative $\frac{\partial
N}{\partial \mu}$ is positive. This gives a clear indication that
the total number of atoms is a monotonic increasing function of
the chemical potential and the soliton solutions to the system of
linearly coupled focusing and defocusing 2D NLS equations are
unstable in the whole domain of their existence. This conclusion
was verified numerically. In figure \ref{FG1} we show the number
of atoms vs. the chemical potential for the 2D-solitons (in all
figures we use the ``scaled number of atoms'' $N$, related to the
actual number of atoms by formula (\ref{EQ18}), and the
dimensionless chemical potential which pertains to the
coupled-mode system (\ref{EQ16})).

For the coupled-mode system with the transverse potential the
behavior of the number of atoms as a function of the chemical
potential is much more interesting. In this case, the total number
of atoms satisfies equations (\ref{EQ55}), (\ref{EQ57}) and
(\ref{EQ59}). Generalizing these to an arbitrary confining
potential, we have:
\begin{equation}
N=\left\{\begin{array}{cc}
|\mu-\mu_\mathrm{bif}|\langle\phi_0^4\rangle^{-1}
\left(1+\frac{\kappa^2}{\mu_2^2}\right)^2\left|1 -
\frac{a\kappa^4}{\mu_2^4}\right|^{-1}
+\mathcal{O}(\mu-\mu_\mathrm{bif})^2, & \varepsilon \ne
\varepsilon_\mathrm{cr},\\
(\mu_\mathrm{bif}-\mu)^{1/2}
\langle\phi_0^6\rangle^{-1/2}\left[\frac{\mu_2^7}{3a^2\kappa^6}
\left(1+\frac{\kappa^2}{\mu_2^2}\right)^3\right]^{1/2}
+\mathcal{O}(\mu-\mu_\mathrm{bif}), & \varepsilon =
\varepsilon_\mathrm{cr},\end{array}\right.
\label{EQ63}\end{equation}
\begin{equation}
N=N_\mathrm{th} + \mu^{-2}(\kappa^2I_1 + I_\mathrm{ext}) +
\mathcal{O}(\mu^{-3}). \label{EQ64}\end{equation} Here $\langle
... \rangle = \int\mathrm{d}^2\vec{\rho}$ and $\phi_0(\rho)$ is
the ground state wave-function of the linear operator $-\nabla^2 +
V_\mathrm{ext}(\rho)$, where $V_\mathrm{ext}(\rho)$ (an even
function) is the confining potential; $I_1$ is given by equation
(\ref{EQ60}), while $I_\mathrm{ext}$ is determined by the
quadratic  term in the Taylor expansion  of the trap
$V_\mathrm{ext}(\rho)$ about $\rho=0$ (for the parabolic trap
$V_\mathrm{ext}= \rho^2$ we have $I_\mathrm{ext} \approx -13.82$).

\begin{figure}[tbp]
\includegraphics{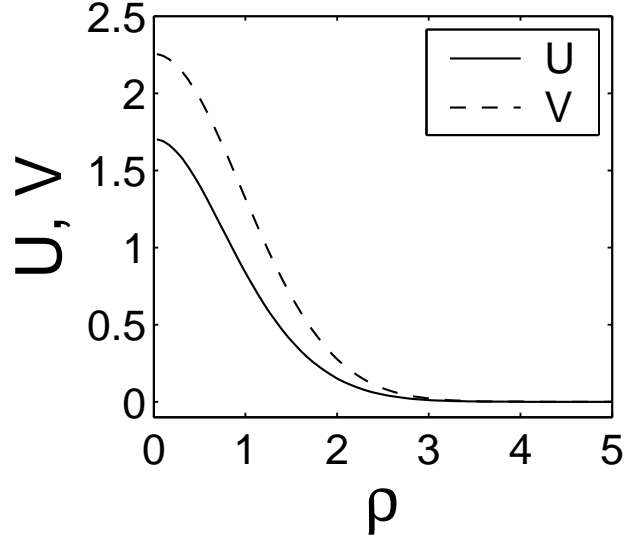}
\caption{\label{FG2} Ground state of the coupled-mode system
(stable with respect to collapse). Here the parameters are $a =
0.005$, $\kappa = 2$, $\varepsilon = -3$, $\mu = -2.31$ and $N =
21.5$. The order parameters $U$ and $V$ are given in the units of
$\sqrt{\Delta}/\ell_\perp$ and the radial length $\rho$  in the
units of $\ell_\perp$.  }
\end{figure}

As shown in the previous section, for $\mu\to-\infty$, the
$u$-component of the solution tends to the Townes soliton, while
the amplitude of the $v$-component tends to zero. Hence, the
solution suffers from the collapse instability in this asymptotic
limit. From equation (\ref{EQ64}) it is seen that the total number
of atoms approaches $N_\mathrm{th}$. However, one cannot conclude
that the collapsing solution is the (unstable) ground state
neither that there are no stable solutions to the coupled-mode
system, which  have the total number of atoms greater than
$N_\mathrm{th}$. In fact, such solutions do exist and give the
ground state of the system, \textit{stable} with respect to
collapse. Figure \ref{FG2} illustrates an example of the stable
ground state  solution with the total number of atoms larger than
$N_\mathrm{th}$, there $a = 0.005$, $\kappa = 2$, $\varepsilon =
-3$ and $N = 21.5$ (it corresponds to a point with $\mu = -2.31$
on the curve of figure \ref{FG4} below). Indeed, equation
(\ref{EQ63}) clearly states that the bifurcating solution is
always stable (by the VK criterion  for $\varepsilon\ge
\varepsilon_{cr}$ and unconditionally otherwise, according to the
discussion in sections \ref{sec3} and \ref{sec5}).

To determine the ground state of the system we have calculated the
energy of the solution numerically. In the coupled-mode
approximation, the energy of the condensate in the double-well
trap is given as
\[
E = \frac{\hbar\omega_\perp
\left(\int\mathrm{d}z|\psi_u|^4\right)^{-1}}{16\pi|a^{(u)}_s|}
\mathcal{H},
\]
\begin{equation}\mathcal{H} = \int\mathrm{d}^2\vec{\rho}\,
\left\{ |\nabla u|^2 + |\nabla v|^2 + \rho^2(|u|^2 +|v|^2)
+\varepsilon |v|^2 - \kappa(uv^* + vu^*) - \frac{|u|^4}{2} +
a\frac{|v|^4}{2} \right\}.
\label{EQ65}\end{equation}
Here $\mathcal{H}$ is the Hamiltonian  of the dimensionless
coupled-mode system.

\begin{figure}[tbp]
\includegraphics{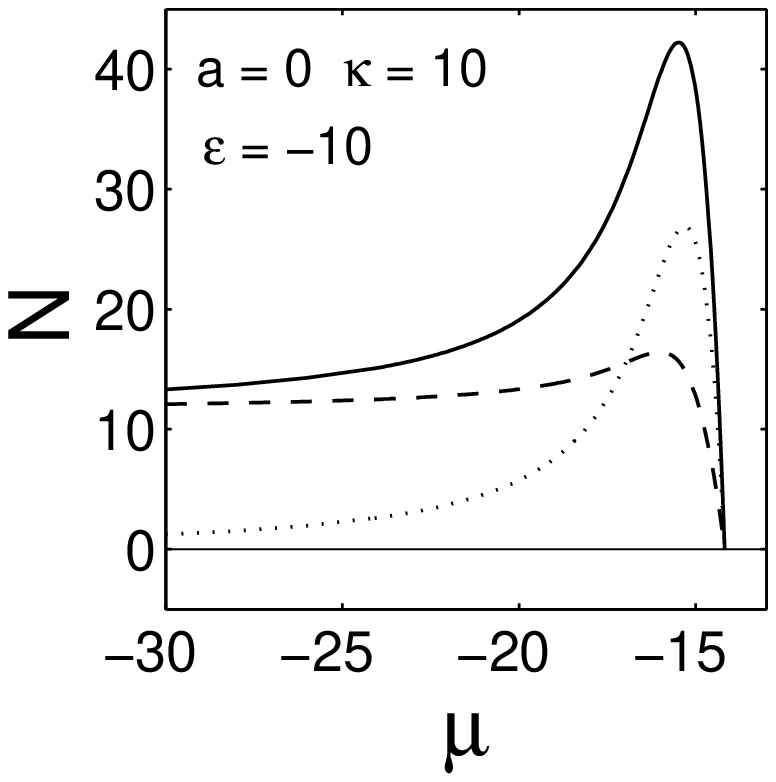}
\caption{\label{FG3} The total number of atoms  vs. the chemical
potential (the solid line) for $a=0$. The dashed and dotted lines
give the number of atoms in the $u$- and $v$-condensates,
respectively. The axes units are as in figure \ref{FG1}. }
\end{figure}

 The existence of stable stationary solutions with large
total number of atoms depends on the system parameters,
principally on the tunnelling coefficient $\kappa$ and the
zero-point energy difference $\varepsilon$. In the asymptotic
limit of weakly coupled equations, i.e., when $\mu\to-\infty$, the
ground state is unstable with respect to collapse, similar as in a
single NLS equation with the external potential. For large values
of $\kappa$ and large negative $\varepsilon$ one can expect the
appearance of a new ground state due to strong quantum tunnelling
through the barrier and competition of the attraction in the
$u$-condensate and the lower zero-point energy for the atoms in
the $v$-condensate. The stable solutions with a large total number
of atoms  were indeed found, for instance, for $a=0$, $\kappa =
10$ and $\varepsilon = -10$; the corresponding dependence of the
total number of atoms on the chemical potential is given in figure
\ref{FG3}. We have checked numerically (by a numerical analysis of
the spectrum of $\Lambda_{1}$ (\ref{EQ29}) from section
\ref{sec3}) that the VK criterion applies. Therefore, the stable
solutions pertain to the region where $\frac{\partial N}{\partial
\mu}<0$,  close to the point of the bifurcation from zero.

Figure \ref{FG3} corresponds to the special case of $a=0$, i.e.
when the $v$-condensate is non-interacting quantum gas. However,
for $a>0$, when the $v$-condensate is repulsive, the curve $N =
N(\mu)$ exhibits similar behavior, see figure \ref{FG4}.

The part of the curve $N = N(\mu)$ with $\frac{\partial
N}{\partial \mu}<0$, corresponding to the stable solutions with  a
large number of atoms, also minimizes the energy when there is a
local maximum with the total number of atoms higher than
$N_\mathrm{th}$. This is illustrated in figure \ref{FG5}, where we
give the equation of state, $\mathcal{H} = \mathcal{H}(N)$ (i.e.,
the energy vs. the total number of atoms in the dimensionless
variables). The equation of state given in this figure is
characteristic for the 2D coupled-mode system for a large region
of values of  the system parameters, when there is a maximum of
the total number of atoms higher than $N_\mathrm{th}$. Otherwise,
the energy takes positive values and the minimum (zero)
corresponds to the collapsing ground state.

Figures \ref{FG3} and \ref{FG4} correspond to the case
$\varepsilon> \varepsilon_\mathrm{cr}$. A new feature appears in
the opposite case, i.e. when $\varepsilon<
\varepsilon_\mathrm{cr}$. Indeed, in this case the effective
equation (see equation (\ref{EQ51}) from section \ref{sec5}) is
defocusing close to the bifurcation point. This fact changes
drastically the dependence of the total number of atoms on
chemical potential: the curve $N = N(\mu)$ enters into the region
of $\mu>\mu_\mathrm{bif}$, i.e. to the right of the bifurcation
point, see figure \ref{FG6}, the inset.  The protruding part of
the curve is analogous to that in a single defocusing 2D NLS
equation with an external potential. The operator $\Lambda_{1}$
appearing in the linear stability analysis of section \ref{sec3}
is positive definite there. Hence,  the solutions corresponding to
this part of the curve $N = N(\mu)$ are unconditionally stable.

Moreover, there is a turning point bifurcation, where
$\frac{\partial N}{\partial \mu} = \infty$ (see the inset in
figure \ref{FG6}). This bifurcation corresponds to the change of
sign of the effective nonlinearity in the coupled-mode system from
negative to positive as one moves upwards along the curve starting
from the bifurcation point. Accordingly, the lowest positive
eigenvalue of the operator $\Lambda_{1,0}$ passes through zero
and becomes negative to the left of the turning point. As there
are no other negative eigenvalues, the VK criterion applies to the
left of the turning point bifurcation.

The equation of state $\mathcal{H} = \mathcal{H}(N)$ corresponding
to figure \ref{FG6}  is similar to that illustrated in figure
\ref{FG5}. Accordingly, the total number of atoms in the ground
state  exceeds $N_\mathrm{th}$ by an order of magnitude in this
case.

Finally, figure \ref{FG7}  illustrates the fact that the maximum
of the total number of atoms achievable by the stationary solution
is determined by the tunnelling coefficient $\kappa$ and the
zero-point energy difference $\varepsilon$ (compare to figure
\ref{FG3}). In this case $\varepsilon<\varepsilon_\mathrm{cr}$
(the turning point bifurcation as well as the protruding part of
the curve are also present, but not visible). The stability is
again determined by the VK criterion, except for the extremely
narrow region before the turning point bifurcation (where the
solution is unconditionally stable). In this case, the total
number of atoms is  always smaller than the threshold
$N_\mathrm{th}$.

As $\mu\to-\infty$  the curves $N=N(\mu)$ and $N_{u,v} =
N_{u,v}(\mu)$ in  figures \ref{FG3} \ref{FG4}, \ref{FG6}, and
\ref{FG7} are similar to that of the ``solitonic'' curve shown in
figure \ref{FG1}. We have confirmed that the  asymptotic solution
for $\mu\to-\infty$ indeed approaches the Townes soliton in its
$u$-component, whereas the $v$-component tends to zero. Thus in
the limit $\mu\to-\infty$ the two condensates are weakly coupled,
similar to the one-dimensional case \cite{PRA}.

The stable ground state in the 2D coupled-mode system, which
corresponds to the part of the curve $N=N(\mu)$  immediately after
the bifurcation point, appears  due to  breaking of the scale
invariance  by the transverse trap. This is reflected in the
estimate of the solution width which is  of the order of the
oscillator length of the trap (see section \ref{sec5}). It is seen
that the larger share of atoms is gathered in the repulsive
$v$-condensate. Hence, such a state is similar to the unusual
bright soliton of the one-dimensional coupled-mode system
\cite{PRA}. However, there is an important difference between the
one-dimensional and the two-dimensional cases: in one spatial
dimension the ground state always corresponds to the weakly
coupled condensates and suffers from the collapse instability for
a large number of atoms.

\section{Conclusion}

Nonlinearity of the Gross-Pitaevsky equation is due to the atomic
interaction which is essential for understanding the properties of
the condensate. Control over the nonlinearity in the
Gross-Pitaevsky equation allows for coupling of two condensates
with the nonlinearities of opposite signs  (the scattering
lengths). This can be realized using a double-well trap with far
separated wells.  In the one-dimensional case, there are stable
bright solitons with almost all atoms gathered in the repulsive
condensate \cite{PRA}. In the two-dimensional case, on the other
hand, the Townes-type solitons in the system are always unstable
due to the fact that the vanishing amplitude 2D-soliton solution
has a finite $l_2$-norm, i.e. the bifurcation from zero always
corresponds to a discontinuity in the dependence of the number of
atoms on the chemical potential. This is quite dissimilar to the
one-dimensional coupled-mode system, where the bifurcation from
zero is continuous except for the boundary case \cite{PRA}.

With the use of a parabolic potential the spatial scale of the
solution is fixed by the oscillator length. This allows for the
stable stationary solutions (though they are not solitons) with
large $l_2$-norms, which represent the ground state of the system.
The ground state is secured from the collapse instability by an
energy barrier. Interestingly, this ground state solution may have
the $l_2$-norm, i.e. the number of atoms of the condensate in a
double-well trap, much higher than the collapse threshold in a
single 2D NLS equation (in some cases, the total number of atoms
exceeds the collapse threshold by an order of magnitude). This is
a new phenomenon, which pertains only to the two-dimensional
coupled-mode system, since in the one-dimensional case, for a
large number of atoms, the ground state corresponds to weakly
coupled condensates, has a large negative chemical potential and
suffers from the collapse instability \cite{PRA}.  A more detailed
study of the properties of the energy barrier for collapse is
relegated to Ref. \cite{EnerBarr}.

\section*{Acknowledgements}
This work was supported by the CNPq and FAPEAL of Brazil.
SBC acknowledges the financial support of Instituto do
Mil\^{e}nio of Quantum Information.

\newpage
\begin{center}\textbf{\large Figures}
\end{center}

\begin{figure}[tbp]
\includegraphics{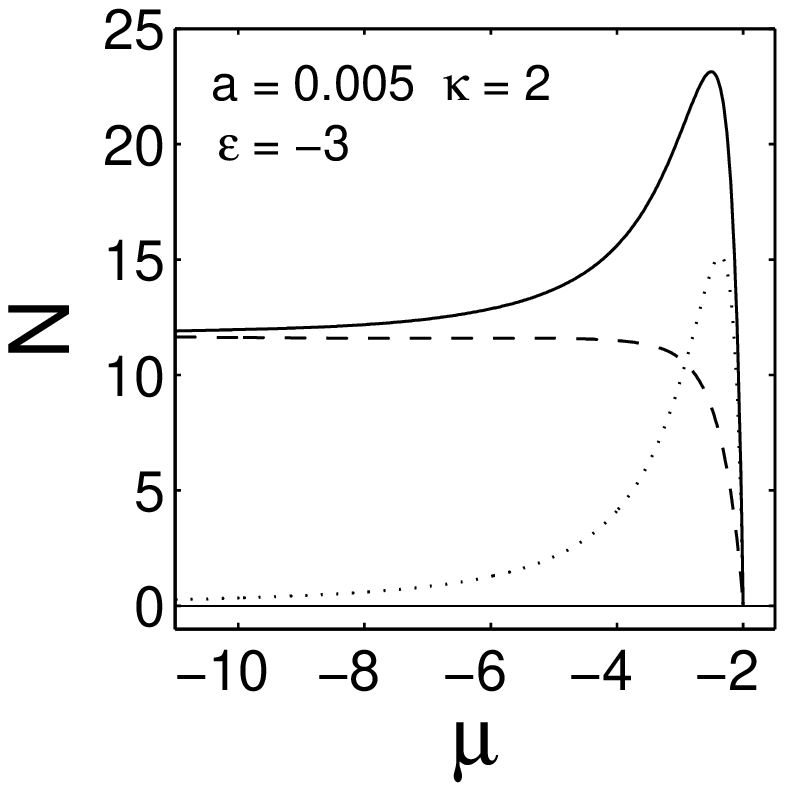}
\caption{\label{FG4} The total number of atoms  vs. the chemical
potential (the solid line) for $a=0.005$. Here $\varepsilon >
\varepsilon_\mathrm{cr}$. The dashed and dotted lines give the
number of atoms in the $u$- and $v$-condensates, respectively. The
axes units are as in figure \ref{FG1}. }
\end{figure}

\begin{figure}[tbp]
\includegraphics{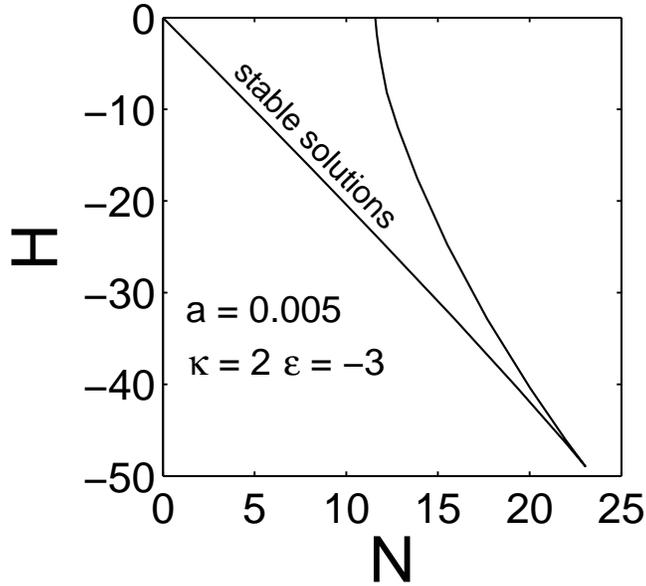}
\caption{\label{FG5} The energy of solutions vs. the total number
of atoms, corresponding to  figure \ref{FG4}. The almost straight
line contains the ground state of the system and corresponds to
the part of the curve $N=N(\mu)$ in figure \ref{FG4} where
$\frac{\partial N}{\partial \mu}<0$. Here the number of atoms is
reduced by the factor $\Delta$ from equation (\ref{EQ18}), whereas
the energy is given in the units of $\hbar\omega_\perp\Delta/2$. }
\end{figure}

\begin{figure}[tbp]
\includegraphics{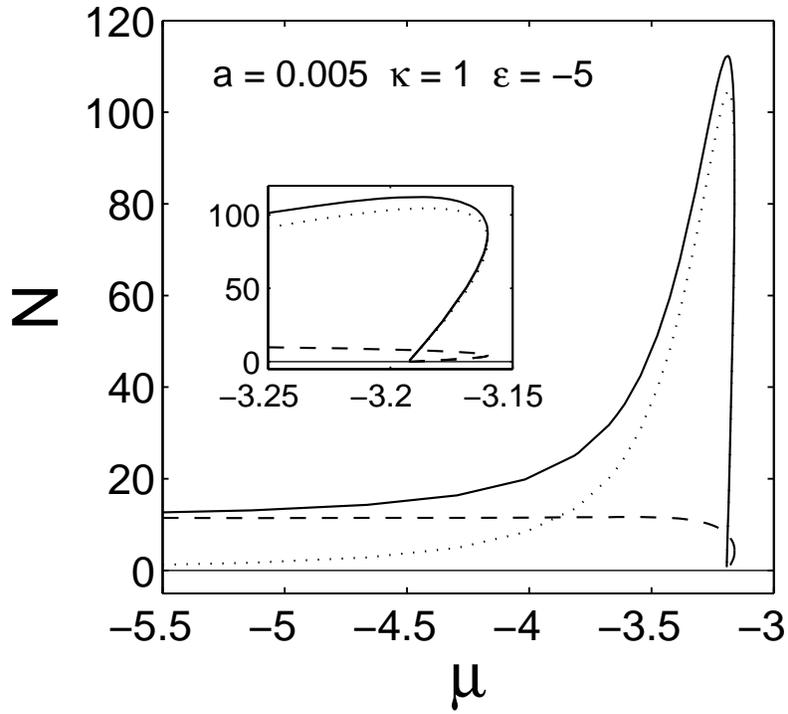}
\caption{\label{FG6} The total number of atoms  vs. the chemical
potential (the solid line). The dashed and dotted lines give the
number of atoms in the $u$- and $v$-condensates, respectively.
Here $\varepsilon < \varepsilon_\mathrm{cr}$. The inset shows a
section of the figure about the bifurcation point. The axes units
are as in figure \ref{FG1}.  }
\end{figure}

\begin{figure}[tbp]
\includegraphics{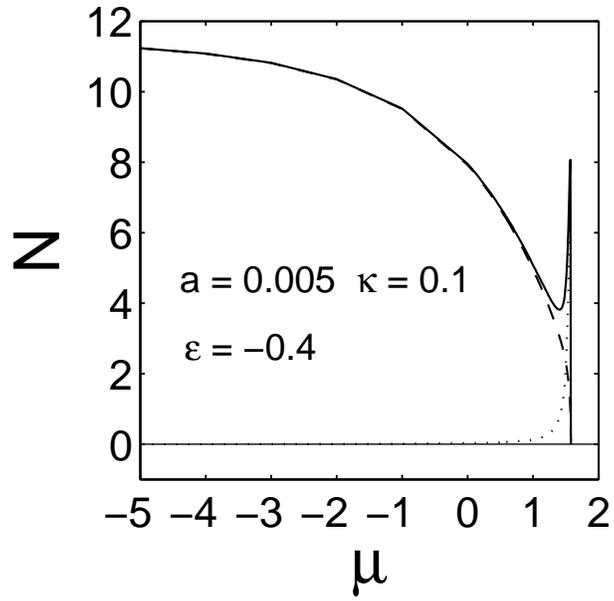}
\caption{\label{FG7} The total number of atoms  vs. the chemical
potential (the solid line) for small values of the tunnelling
coefficient and energy difference.  Here $\varepsilon
<\varepsilon_\mathrm{cr}$. The dashed and dotted lines give the
number of atoms in the $u$- and $v$-condensates, respectively. The
axes units are as in figure \ref{FG1}.  }
\end{figure}

\end{document}